\documentclass[conference]{IEEEtran}
\usepackage{cite}
\usepackage{amsmath,amssymb,amsfonts}
\usepackage{algorithmic}
\usepackage{booktabs}
\usepackage{graphicx}
\usepackage{color}
\usepackage{textcomp}
\usepackage{listings}
\usepackage{siunitx}
\usepackage{tikz}
\usepackage{url}
\usepackage{alltt}

\def\BibTeX{{\rm B\kern-.05em{\sc i\kern-.025em b}\kern-.08em
    T\kern-.1667em\lower.7ex\hbox{E}\kern-.125emX}}

\title{Semantic Identifiers and DNS Names for IoT
}

\graphicspath{{./figs/}}
\DeclareGraphicsExtensions{.pdf}

\author{\IEEEauthorblockN{Simon Fernandez\IEEEauthorrefmark{1}, Michele
	Amoretti\IEEEauthorrefmark{5}, Fabrizio Restori\IEEEauthorrefmark{5},
        Maciej Korczyński\IEEEauthorrefmark{1}, Andrzej Duda\IEEEauthorrefmark{1}}
		\IEEEauthorblockA{\IEEEauthorrefmark{1}Univ. Grenoble Alpes, CNRS, Grenoble
		INP, LIG, F-38000 Grenoble, France}
		\IEEEauthorblockA{\IEEEauthorrefmark{5}Department of Engineering and
		Architecture, University of Parma, Italy}
}

\newcommand{\ttt}[1]{\textbf{\texttt{\small{#1}}}}

\begin{document}

\maketitle

%\todo{Remove this table of contents. Just for debug purpose.}
%\tableofcontents

\begin{abstract}
In this paper, we propose a scheme for representing semantic metadata of IoT
devices in compact identifiers and DNS names to enable simple discovery and search
with standard DNS servers.
Our scheme defines a binary identifier as a sequence of bits: a Context to use
and several bits of fields corresponding to semantic properties specific to the
Context. 
The bit string is then encoded as \ttt{base32} characters and
registered in DNS. 
Furthermore, we use the compact semantic DNS names to offer support for search
and discovery. We propose to take advantage of the DNS system 
as the basic functionality for querying and discovery of semantic
properties related to IoT devices. 

% The encoding scheme of semantic metadata
% structures the DNS names similarly to IP prefixes: a longer prefix
% represents more specific semantic information and shortening a prefix corresponds to more
% general information, thus allowing for some range/region or property queries.

We have defined three specific Contexts for hierarchical semantic properties as
well as logical and geographical locations. 
For this last part, we have developed two prototypes for managing geo-identifiers
in LoRa networks, one based on Node and the Redis in-memory
database, the other one based on the CoreDNS server.  
\end{abstract}


%%%%%%%%%%%%%%%%%%%%%%%%%%%%%%%%%%%%%%%%%%%%%%%%%%%%%%
\section{Introduction}

Many IoT applications require the knowledge about the various properties of IoT
devices that provide some data about the physical world and can act upon the
environment. The properties may for instance include the information on:

\begin{itemize}
  \item type and unit of data, (e.g., temperature in °C),
  \item resolution, frequency of data (e.g., $512\times512$ pixels every hour),
  \item possible actions performed by the device (e.g., switch on),
  \item raised alarms (e.g., overheating),
  \item geographic location of the device (e.g., $(+28.61, -80.61)$ WGS84/GPS coordinates),
  \item logical location of the device (e.g., Room 235 on Floor 14).
\end{itemize} 

Several initiatives aimed at expressing and structuring this kind of IoT and M2M
metadata: Sensor Markup Language (SenML)~\cite{senml}, IPSO Alliance
Framework~\cite{IPSO}, and oneM2M Base ontology~\cite{AlayaMMD15}. 
The World Wide Web Consortium (W3C) schemes for the semantic Web such as
RDF,\footnote{\url{http://www.w3.org/RDF}}
OWL,\footnote{\url{http://www.w3.org/TR/owl-ref}}
SPARQL\footnote{\url{http://www.w3.org/TR/sparql11-query/}} also allow 
understanding and discovery of IoT data.
For expressing specific IoT semantics, W3C proposed a Semantic Sensor Network (SSN)
ontology\footnote{\url{https://www.w3.org/2005/Incubator/ssn/ssnx/ssn}} that
allows the description of sensors and their characteristics addressing the
issue of interoperability of metadata annotations.

The Web of Things initiative of W3C\footnote{\url{https://www.w3.org/WoT}} aims at unifying 
IoT with digital twins for sensors,
actuators, and information services exposed to applications as
local objects with \emph{properties}, \emph{actions}, and \emph{events}.
W3C Thing Description (TD)\footnote{\url{https://www.w3.org/TR/wot-thing-description}}
expressed in JSON-LD\footnote{\url{https://www.w3.org/TR/json-ld}} covers the
behavior, interaction affordances, data schema, security configuration, and
protocol bindings. 

Thing Description allows for attaching rich semantic metadata to IoT devices,
however, this format is oriented towards processing by non-constrained
applications running for example in the Cloud or in the Edge to become the base
for sophisticated discovery and search services offered on Web servers for IoT
users and applications. 
However, we can notice that discovery and search based on semantic metadata also
happens in constrained IoT environments---an IoT device needs to
discover other devices and choose the right one for further communication or
collaboration.
In this case, semantic metadata of IoT devices need to be
encoded in a highly compact way to reduce the overhead in usually bandwidth
limited networks. 

In this paper, we propose a scheme for representing semantic metadata of IoT
devices in compact identifiers or names to enable simple discovery and search
with standard DNS servers. 
The idea of the scheme is inspired by the Static Context Header Compression
(SCHC)\footnote{\url{https://tools.ietf.org/html/rfc8724}} approach to IP header
compression.
In SCHC, two devices that exchange IP packets compress headers based on
pre-established contexts. Instead of a full header, a device inserts the
information about the context to use and some short information required to
reconstruct the header. In this way, a 40 byte IPv6 header can be compressed
down to just a few bytes. 
Our scheme defines a binary identifier as a sequence of bits composed
of a Context to use and several fields corresponding to semantic properties specific to the
Context. 
The bit string is then encoded as \ttt{base32} characters and
registered in DNS. Thus, the DNS name encodes in a compact form the semantic
metadata of an IoT device.  

We define several Contexts of identifiers expressing different semantic
metadata to fit the most popular device characteristics (other can also be defined): 
\begin{enumerate}
  \item hierarchical semantic properties, 
  \item logical location of the device,
  \item geographic location of the device.
\end{enumerate}
The first one corresponds to the structured representation of 
the attributes of Thing Description and two others cover the geographical
information about an IoT device.
We instantiate the scheme for encoding geographic location in case of LoRa
networks and show how to construct a 64 bit geo-identifier of LoRa devices. 

Furthermore, we use the compact semantic DNS names to offer support for search
and discovery. In constrained environments, providing full-fledged database
search functionality may be difficult. Instead, we propose to take advantage of
the DNS system as the basic functionality for querying and discovering the semantic
properties related to IoT devices. Our encoding scheme of semantic metadata
structures the DNS names similarly to IP prefixes: a longer prefix
represents more specific information and shortening a prefix corresponds to more
general information, thus allowing for some range or extended topic queries. 
For instance, if the name represents a geographical location, a longer name
represents a smaller area and a shorter name corresponds to a larger zone that
encompasses the smaller area designated by the longer name.
Finally, we describe two prototypes supporting DNS queries on geo-identifiers. 

Querying DNS based on semantic names can bring interesting features to many IoT
applications: finding devices corresponding to a given property, placement on a
map of all sensors belonging to a given application, sending commands to the
devices in a chosen region, or gathering data from chosen devices based on their
geographical location.

The paper makes the following contributions:
\begin{enumerate}
  \item we define a scheme based on Contexts for compact encoding of different
    types of metadata in DNS names,
  \item we take advantage of geohashes to instantiate the scheme for encoding geographic location,
  \item we propose a means for simple and minimal discovery of IoT devices and
    searching for their characteristics based on standard DNS functions,
  \item we explore an idea of using DNS to store and publish IoT data,
  \item we validate the proposed schemes with preliminary prototypes
    supporting DNS queries on geo-identifiers. 
\end{enumerate}

% The rest of the paper is organized as follows. 
% We discuss previous work in Section~\ref{sec:related}. 
% Section~\ref{sec:encoding} describes the principles of compact encoding of
% IoT metadata and define specific contexts for hierarchical
% properties and logical location. We define the
% identifiers and names for geographic location in
% Section~\ref{sec:geo-encoding}. 
% Section~\ref{sec:impl} shows how we can take advantage of DNS Service Discovery
% for enabling semantic queries on names and Section~\ref{sec:proto}  describes our prototypes.
% Section~\ref{sec:dns-data} presents the idea of using DNS to store and publish IoT data. 
% Finally, we draw some conclusions in Section~\ref{sec:conclusion}.

\section{Related Work}
\label{sec:related}

We briefly review previous work related to expressing semantic properties of IoT
devices and compact encoding of geographical location.

\subsection{Semantic Properties of IoT Devices}

As mentioned in the introduction, several initiatives considered the problem of
expressing metadata of IoT devices and M2M communications: Sensor Markup
Language (SenML)~\cite{senml}, IPSO Alliance 
Framework~\cite{IPSO}, and oneM2M Base ontology~\cite{AlayaMMD15}. 
Kovacs et al. proposed a system architecture for achieving worldwide semantic
interoperability with oneM2M~\cite{7785888}. 
The Semantic Sensor Network (SSN) ontology allows the description of sensors and
their characteristics~\cite{HallerJCLTPLGAS19}.

An important initiative of W3C aimed at creating the semantic Web of
Things~\cite{PfistererRBKMTHKPHKLPR11} to enable unambiguous exchange of IoT
data with shared meaning.  
Novo and Di Francesco~\cite{NovoF20} discussed solutions
that extend the Web of Things architecture to achieve a higher level of semantic
interoperability for the Internet of Things. 
Nevertheless, many proposed approaches do not address the constraints of IoT
devices that do not match the size and the form of semantic descriptions usually
developed in the traditional W3C setting. 
For instance, Novo and Di Francesco~\cite{NovoF20} reported performance results coming
from a testbed composed of two computers connected to an 802.11 network. 

We proposed DINAS~\cite{AmorettiAFRD17}, a scheme based on Bloom filters for
creating compact names from node descriptions and a service discovery protocol
for short-range IoT networks running RPL.
Other work emphasizes the importance of DNS for IoT~\cite{9133283}.

%We review below different ways of encoding geographical coordinates. 

\subsection{WGS84 aka GPS}

WGS84 is a common format for encoding geographical coordinates used in GPS,
composed of two numbers in degrees of the form \ttt{ddd.ddddddd}, where \ttt{d} stands for a degree
digit. Degrees are expressed as numbers between $-180$ and $+180$ for longitude,
and a number between $-90$ and $+90$ for latitude (locations to the west and to
the south are negative), e.g., $(+28.61, -80.61)$ corresponds to the location of
the Cape Canaveral Space Center.

Expressing a given geographical location is always done with a given precision,
and when decoding a position, all methods return the center of the square
representing all possible positions.
For example, if we decode $(28^\circ\mathrm{N}, 80^\circ\mathrm{W})$, we know the position
is in the square between $(28^\circ\mathrm{N}, 80^\circ\mathrm{W})$ and $(29^\circ\mathrm{N},
81^\circ\mathrm{W})$, and we will return $(28.5^\circ\mathrm{N}, 80.5^\circ\mathrm{W})$ to
minimize the error.

Table~\ref{tab:size} represents the longitudinal resolution at the equator and
at a latitude of $45^\circ\mathrm{N}/\mathrm{S}$ with an increasing number of
decimal figures and the corresponding number of bits to represent them.
The idea is to relate the size of a region to the number of bits used for
representing a given geographical coordinate and thus relate the size of a
region to the size of an identifier. We can observe that 8 decimal figures
encoded on 26 bits are sufficient to represent the location at the precision of
around 1\,m.

\begin{table}[t]
  \centering
  \caption{Longitudinal decimal degree precision}
  \begin{tabular}{ccr@{.}lr@{.}l}
    \toprule
    \bf \# of figures & \bf \# of bits & \multicolumn{2}{c}{\bf Equator} &
    \multicolumn{2}{c}{\bf $45^\circ \mathrm{N}/\mathrm{S}$} \\
    \midrule
    3 & 9  & 111 & 3200 km & 78  & 710 km  \\
    4 & 12 & 11  & 1320 km & 7   & 871 km  \\
    5 & 16 & 1   & 1132 km & 787 & 100 m  \\
    6 & 19 & 111 & 3200 m  & 78  & 710 m  \\
    7 & 22 & 11  & 1320 m  & 7   & 871 m  \\
    8 & 26 & 1   & 1132 m  & 787 & 100 mm \\
    \bottomrule
  \end{tabular}\label{tab:size}
\end{table}

\subsection{Geoprefixes, Geohashes, Plus Codes}

In previous work, we defined the notion of a {\em geoprefix}
for IPv6 networks~\cite{brunisholzDataTweetUsercentricGeocentric2016}: the
location of each device is encoded in its IPv6 multicast address and an
application can send a packet to all devices corresponding to a given prefix
representing a geographic area (a geocast).

Niemeyer~\cite{geohash} proposed {\em geohash}, an encoding of WGS84 coordinates
based on Morton codes~\cite{morton1966computer} that computes a 1-dimensional
value from the 2-dimensional GPS coordinates by interleaving the binary
representations of the coordinates and further represented as ASCII characters.

In this method, to encode a given location, we proceed by a dichotomy. Starting
with the full interval (${[-180; +180]}$ for longitude, ${[-90; +90]}$
for latitude), we split the interval in two (${[-90; 0]}$ and ${[0; +90]}$
for latitude), then, if the location is in the higher half, we add bit
\ttt{1} to the encoding of the coordinate, or else, we
add bit \ttt{0}, and we repeat the operation with the
new interval, building the encoding bit by bit, until we reach the desired
precision. When decoding, once the last bit is reached, the decoded location
is at the center of the remaining interval (for latitude, if we have only one
bit with value \ttt{1}, we would decode that the latitude
is ${+45}$). With this method, each additional bit halves the size of the
interval.

Once both latitude and longitude are represented this way, their binary codes
are intermingled to produce a unique value: odd bits represent latitude and
even bits represent longitude as presented in Table~\ref{fig:geohash_mix}. For
example, the resulting encoding of latitude \ttt{1011
1100 1001} and longitude \ttt{0111 1100 0000} is
\ttt{0110 1111 1111 0000 1000 0010}.

\begin{table}[t]
  \centering
  \caption{Combining latitude and longitude encoded in a unique binary value}
  \begin{tabular}{rl}
    Long.    & \ttt{0.1.1.1.1.1.0.0.0.0.0.0.0} \\
    Lat.     & \ttt{ .1.0.1.1.1.1.0.0.1.0.0.1.} \\
    \midrule
    Result   & \ttt{0110111111110000010000010} \\
  \end{tabular}\label{fig:geohash_mix}
\end{table}

\begin{table}[t]
  \centering
  \caption{Longitudinal decimal degree precision and the size of a geohash}
  \begin{tabular}{p{0.7cm}p{0.7cm}p{0.7cm}rrr}
    \toprule
length & lat bits & lng bits & lat error & lng error &
error\\
    \midrule
1    & 2& 3    & $\pm$ 23° & $\pm$ 23°& $\pm$ 2500 km\\
2    & 5& 5    & $\pm$ 2.8° & $\pm$ 5.6°& $\pm$ 630 km    \\
3    & 7& 8    & $\pm$ 0.70° & $\pm$ 0.70°& $\pm$ 78 km \\
4    & 10    & 10   & $\pm$ 0.087° & $\pm$ 0.18°& $\pm$ 20 km \\
5    & 12    & 13   & $\pm$ 0.022° & $\pm$ 0.022°& $\pm$ 2.4 km\\
6    & 15    & 15   & $\pm$ 0.0027° & $\pm$ 0.0055°& $\pm$ 610 m\\
7    & 17    & 18   & $\pm$ 0.00068° & $\pm$ 0.00068°& $\pm$ 76 m    \\
8    & 20    & 20   & $\pm$ 0.000085° & $\pm$ 0.00017°& $\pm$ 19 m   \\
9    & 22    & 23   &&    & \\
10   & 25    & 25   &&    & $\pm$ 59 cm  \\
11   & 27    & 28   &&    & \\
12   & 30    & 30   &&    & $\pm$ 1.84 cm\\
    \bottomrule
  \end{tabular}\label{tab:size2}
\end{table}

{\em Geohash-36},\footnote{\url{http://en.wikipedia.org/wiki/Geohash-36}} originally developed for compression of world
coordinate data, divides the area into 36 squares and generates a full character
describing which sub-square contains the position. 

Google Maps uses {\em Plus Codes}~\cite{pluscodes,pluscodes2} made up of a
sequence of digits chosen from a set of $20$. 
The digits in the code alternate between latitude and longitude. The first four
digits describe a one degree latitude by one degree longitude area, aligned on
degrees. 
A Plus Code is 10 characters long with a plus sign before
the last two:
\begin{enumerate}
\item
The first four characters are the area code describing a region of roughly 100
$\times$ 100 kilometers.

\item
The last six characters are the local code, describing the neighborhood and the
building, an area of roughly 14 $\times$ 14 meters.
\end{enumerate}

As an example, let us consider the Parliament Buildings in Nairobi, Kenya
located at the \ttt{6GCRPR6C+24} plus code: \ttt{6GCR} is the area from \ttt{2S
36E} to \ttt{1S 37E}. \ttt{PR6C+24} is a 14-meter wide by 14-meter high area
within \ttt{6GCR}. The \ttt{+} character is used after eight digits, to break
the code up into two parts and to distinguish codes from postal codes.

\section{Compact Encoding of IoT Metadata}\label{sec:encoding}

\begin{figure}[t!]
\begin{centering}
\includegraphics[width=\linewidth]{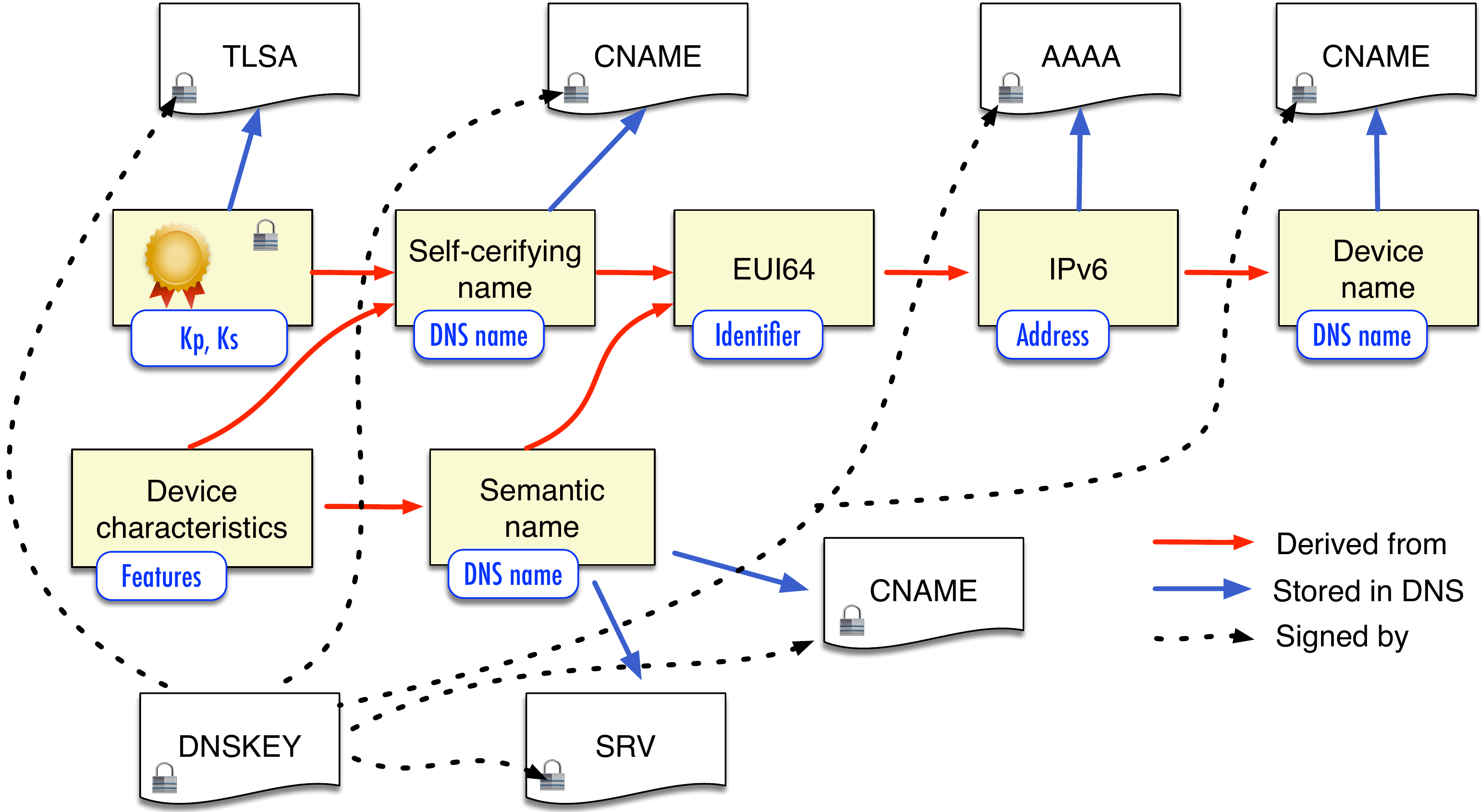}
\caption{General scheme for identifiers and names.} 
\label{fig:data}
\end{centering}
\end{figure}

The main objective of this paper is to design a scheme for encoding semantic
properties in DNS names so that IoT devices could discover relevant nodes using
with DNS name resolution. 
Figure~\ref{fig:data} gives an example of how it can be done in the context of LoRa
devices. Note that DNSSEC guarantees the information integrity.

We propose to assign \emph{self-certifying names} to IoT devices: the name derives
from a public key to enable secure establishing of the identity of a device
without relying on an external PKI infrastructure.  
The self-certifying name is constructed as a hash of public key $K_p$ similarly to Bitcoin addresses:

\[A = ripemd160(sha256(K_p))\]

then $A$ is encoded with \ttt{base32} (20 characters) giving the
DNS name $N$. 
\ttt{base32} encoding represents a binary string with
\ttt{0-9} digits and some lower case letters (excluding characters hard to
distinguish like \ttt{i}, \ttt{l}, \ttt{o}). 
We cannot use
\ttt{base58check} like in Bitcoin because of capital/low case letters in
\ttt{base58check} (DNS names do not distinguish between capital/low case
letters). 
The nice feature of this scheme is that devices can check whether a public key
from the TLSA DNS record corresponds to the name and if authentication is
enforced (signature with the private key $K_s$) to be sure a device
communicates with the right peer. 

Then, we can derive an 8 byte \ttt{EUI64} identifier from $A$ with SHA-3($A$).
\ttt{EUI64} identifiers are required in some networks like LoRa---we can obtain
the LoRa
\ttt{DevEUI} identifier derived from $K_p$ and then use it to
construct an IPv6 address. We will also show below that the
\ttt{DevEUI} of a LoRa device can represent its geographical location. 

In addition to the self-certifying name, we define other names (DNS aliases)
that represent device properties encoded in a compact
way. 
Moreover, we want the encoding scheme to take advantage of some discovery
functionalities of DNS by requiring that a name is structured as an IP
prefix---smaller prefix means a more general query. 

\subsection{Encoding Hierarchical Semantic Properties}
\label{sec:properties}

\begin{figure}[t!]
\begin{centering}
\includegraphics[width=\linewidth]{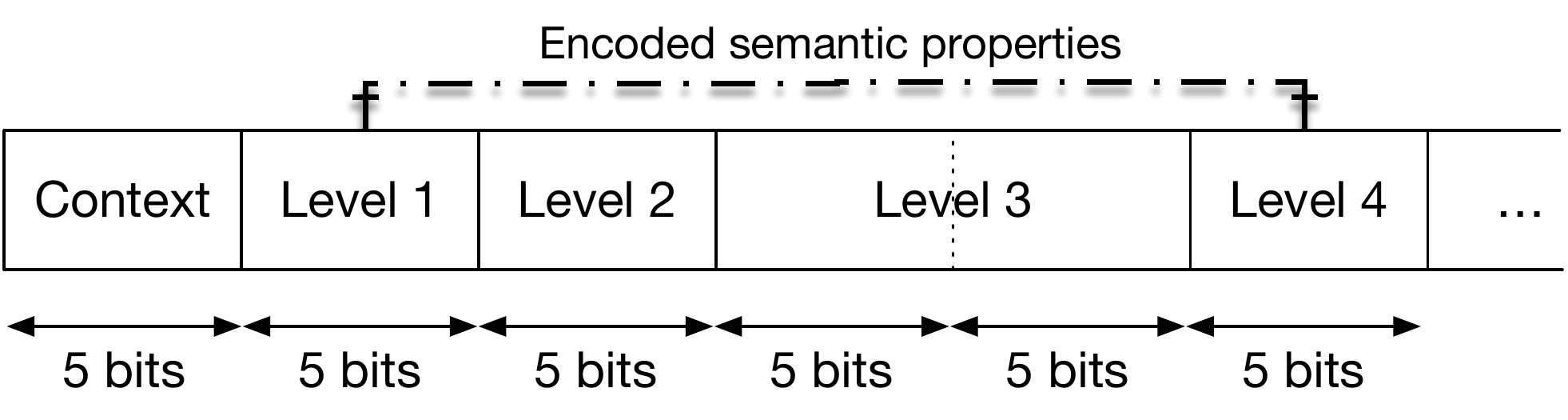}
\caption{Structure of a binary semantic identifier (fields of 5 bits or a multiple of 5 bits).} 
\label{fig:struct}
\end{centering}
\end{figure}

\begin{figure}[t!]
\begin{centering}
\includegraphics[width=\linewidth]{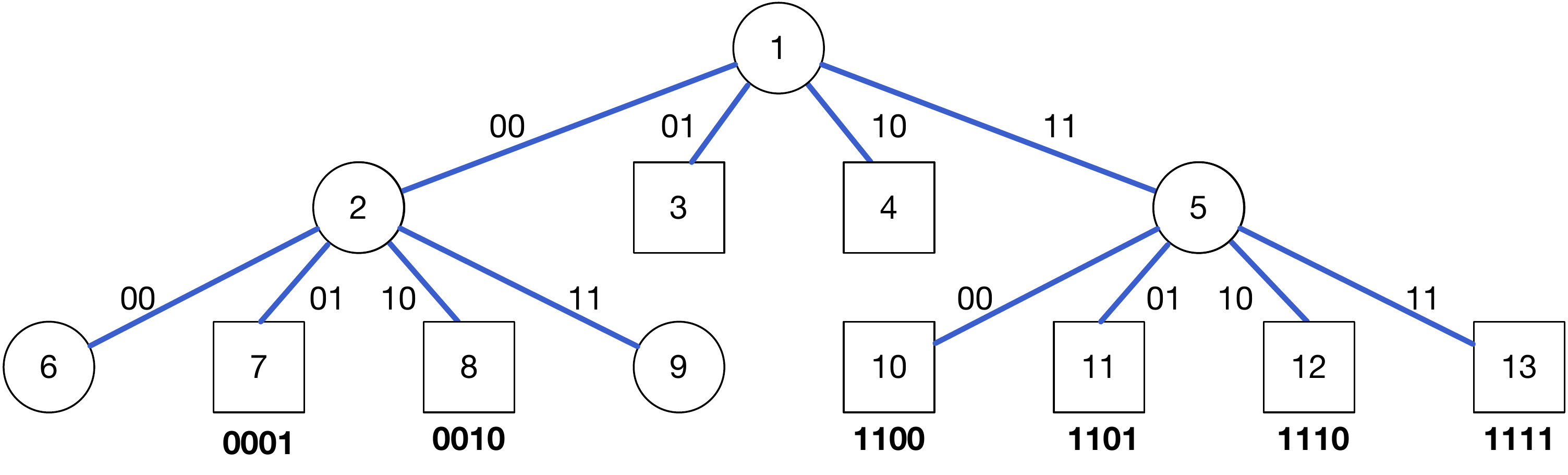}
\caption{Semantic attributes encoded as a quadtree.} 
\label{fig:tree}
\end{centering}
%\vspace{-2ex}
\end{figure}

Figure~\ref{fig:struct} presents the structure of an identifier encoding a semantic tree of several
levels presented below.
A Context defines how to interpret the encoded semantic properties. 
The first type of Contexts we define is a hierarchical representation of properties in a
form of a semantic tree with leaves corresponding to properties (see example in Figure~\ref{fig:tree}). 
The identifier encoding the semantic properties is the binary code generated
when traversing the tree. 
To be able to express the binary identifier of a level with one or more
\ttt{base32} letters, each field is 5 bits or a multiple of 5 bits (e.g., the
Level 3 field composed of 10 bits). 
In this way, we avoid problems of dealing with padding if the size of the binary
identifier is not a multiple of 5 bits. 
Note that the structure of a given identifier (the size of each field) is
defined by a given Context. 

Figure~\ref{fig:tree} presents an example of a semantic quadtree of
degree $n=4$ for simplicity. 
In fact, we use $n=32$ for 5 bits assigned
to each level of the tree, which gives the possibility of representing
$32^2=1024$ attributes at Level 2 or more, e.g., $32,768$ at Level 3 with the
field of 10 bits.
We may have trees with different numbers of nodes at each level depending on the
need of representing more or less properties (e.g., $n=1024$, if we assign 10 bits
to a level), the only limitation being that the
level degree should be a power of $2$. 
Circles correspond to non-terminal nodes and squares to 
leaves. 

The position in the tree determines the code of a property, for instance, the
property at leaf $11$ has the code of \ttt{1101} corresponding
to the traversal of the \ttt{11} branch and then, the
\ttt{01} one. 
One identifier (so one DNS name) corresponds to one leaf in the tree. 

The encoding scheme structures the DNS names similarly to IP prefixes: a longer
prefix represents more specific information and shortening a prefix corresponds
to more general information, thus allowing for some range or property queries,
e.g., the shorter code of \ttt{11} represents all properties with \ttt{1100}
\ttt{1101} \ttt{1110} \ttt{1111} identifiers. 

Below, we present an example of creating a semantic name for a
temperature sensor with metadata expressed in the
following W3C Thing Description:\footnote{\url{https://www.w3.org/TR/wot-thing-description}}

%\vspace{1em}

\begin{footnotesize}
\begin{alltt}
\textbf{"@context": [}
\textbf{    "https://www.w3.org/2019/wot/td/v1",}
\textbf{...],}
\textbf{"@type": "saref:TemperatureSensor",}
\textbf{...}
\textbf{"properties": \textbraceleft}
\textbf{    "temperature": \textbraceleft}
\textbf{        "description": "Weather Station Temperature",}
\textbf{        "type": "number",}
\textbf{        "minimum": -32.5,}
\textbf{        "maximum": 55.2,}
\textbf{        "unit": "om:degree_Celsius",}
\textbf{        "forms": [...]}
\textbf{    \textbraceright,}
\textbf{    ...}
\textbf{\textbraceright,}
\textbf{...}
\end{alltt}
\end{footnotesize}

\noindent
where some attributes refer to external vocabularies such as
SAREF (Smart Appliances REFerence
Ontology)\footnote{\url{https://ontology.tno.nl/saref/}} and
OM
(Ontology of Units of Measure)\footnote{\url{http://www.ontology-of-units-of-measure.org/page/om-2}}.

The encoded binary identifier is composed of the following fields (\ttt{base32}
encoding in parenthesis\footnote{In all our encoding examples, we use the
  \textbf{\texttt{\footnotesize base32}} version defined for geohashes that encode geographical
  coordinates instead of the version described by RFC 4648,
  \url{https://tools.ietf.org/html/rfc4648}}.):

\begin{itemize}
  \item \ttt{00001} — Context-1 (\ttt{1}) 
  \item \ttt{01100} — properties (\ttt{d})
  \item \ttt{00001} — temperature (\ttt{1})
  \item \ttt{00101} — unit (\ttt{5})
  \item \ttt{00010} — degree\_Celsius (\ttt{2})
\end{itemize}

We assume that Context-1 corresponds to the semantic tree generated according to
the TD context with 5 bits per level, \ttt{properties} is the 12th attribute,
\ttt{temperature} the 1st one, and the value of the unit \ttt{degree\_Celsius}
is the 2nd possible value. 
The binary identifier of \ttt{00001 01100 00001 00101 00010} results in the
\ttt{1d152} DNS name. 

\subsection{Encoding Logical Location}
\label{sec:logical}

In many use cases, an IoT application may benefit from metadata about
localization in a logical form.  
For instance, when defining group communication for the Constrained Application
Protocol (CoAP), RFC~7390\footnote{\url{https://tools.ietf.org/html/rfc7390}} considered a building
control application that wants to send packets to a group of nodes represented
by the following name:

%\noindent
\ttt{all.bu036.floor1.west.bldg6.example.com}.

\noindent Logically, the group corresponds to \emph{"all nodes in office bu036,
  floor 1, west wing, building 6"}. 
Such hierarchical groups of fully qualified domain naming (and scoping) provide a logical
description of places that may complement other precise geographical information
that we consider in the next section. 

We can observe that there is an inclusion relationship between elements of the
description: office \ttt{bu036} is on the \ttt{floor 1}, at the \ttt{west} wing of 
\ttt{building 6}.
A specific Context can represent the inclusion relationship in a semantic tree,
so we can express a given logical location with a binary identifier and an encoded
DNS name. 

Let us assume that we have up to 32 buildings, 32 floors, and 1024 rooms. 
To encode the location of Room 214 on Floor 19 in Building 7, we define 
the binary identifier composed of the following fields (\ttt{base32}
encoding in parenthesis):

\begin{itemize}
  \item \ttt{00010} - Context-2 (\ttt{2}) 

  \item \ttt{00111} - Building 7 (\ttt{7})

  \item \ttt{10011} - Floor 19 (\ttt{m})

  \item \ttt{01011} - Room 376 - first part (\ttt{c})

  \item \ttt{11000} - Room 376 - second part (\ttt{s})
\end{itemize}

We assume that Context-2 corresponds to the semantic tree with 5 bits at the 1st
and 2nd levels, and 10 bits at the 3rd level as in Figure~\ref{fig:struct}. 
The binary identifier \ttt{00010 00111 10011 01011 11000} results in the
\ttt{27mcs} DNS name.

\section{Encoding Geographic Location}
\label{sec:geo-encoding}

Many IoT applications require precise information on the geographical location
of IoT devices---when a sensor provides some measurement data, one of the most
important additional information is the
localization of the data source, usually stored as metadata.
We can use GPS for localization, however, adding GPS to an IoT
device increases its cost and energy consumption, which may make their cost prohibitive for
many large scale IoT applications.

\begin{figure}[t!]
\begin{centering}
\includegraphics[width=0.95\linewidth]{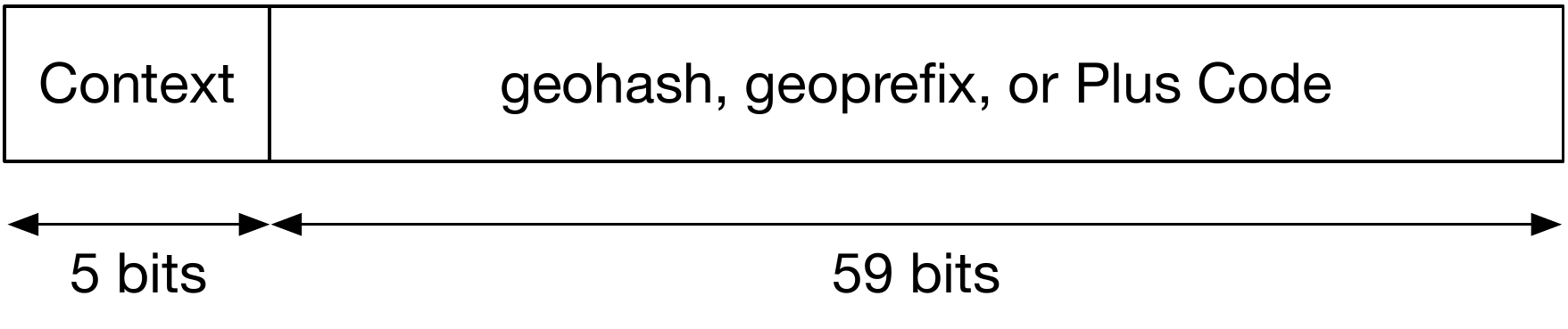}
\caption{Structure of a geo-identifier on 64 bits.} 
\label{fig:struct2}
\end{centering}
%\vspace{-0.5em}
\end{figure}

We take the example of LoRa networks to consider the problem of
representing geographical locations in identifiers and DNS names. 
We propose a scheme to define the \emph{geo-identifier} of
a LoRaWAN device in 
a way that encodes its geographical location.
Figure~\ref{fig:struct2} presents its structure with two fields: 5
bits for the Context and 59 bits for encoding geographical coordinates of
different forms. 
The Context gives the information about the type of encoding. 

LoRaWAN defines \ttt{DevEUI}, a unique 64 bit identifier in
the IEEE EUI-64~\cite{EUI-64} configured on a device.
In the activation process of the device, it obtains \ttt{DevAddr}, a 32 bit
identifier in the current network allocated by the Network Server.
We propose to use a geo-identifier as \ttt{DevEUI}, store \ttt{DevEUI} as a DNS
name, and provide a lookup service-based DNS service discovery that returns
names corresponding to a geographical region. 

Then, we can encode a binary string in the geohash variant of \ttt{base32} giving us a name. 
For instance, 
the binary value of \ttt{01101 11111 11000 00100 00010} (25 bits) results in the
\ttt{ezs42} geohash composed of 5 characters, the representation having a
location error of $\pm$ 2.4 km.

With 59 bits for encoding the latitude and the longitude, a geoprefix or a
geohash result in a resolution of a few cm. 
Table~\ref{tab:size2} presents the size of the \ttt{base32} encoded
geohash, the number of bits 
representing longitude and latitude, and the precision of the decoded value.

Geohashes offer interesting features: i) similar geohashes represent nearby
positions and ii) a longer geohash represents a smaller area and
shortening it reduces the precision of both its coordinates to represent a
larger region.

\begin{table}[t]
  \centering
  \caption{Practical example of a geohash}
  \begin{tabular}{lll}
    geohash               & Latitude   & Longitude \\
    \midrule
    \texttt{dr5r7p4rx6kz} & 40.689167 & -74.044444 \\
    \texttt{\textbf{dr5r}7p4}     &  40.69   & -74.04 \\
    \texttt{\textbf{dr5r}111}     &  40.61    & -74.13 \\
  \end{tabular}\label{tab:pref_example}
%\vspace{-1em}
\end{table}

Table~\ref{tab:pref_example} presents a practical example of the prefix property
of a geohash. In
the table, the second geohash is a prefix of the first one, so it is more
precise and is contained in the second one. The second and the third geohash
have a common prefix, so they are in the same region (easily computable with
the common prefix) but do not overlap.

%Elasticsearch, MongoDB, HBase, Redis, and Accumulo successfully used geohashes
%to implement proximity searches.

% There is also 64-bit variant of geohash~\cite{geohash-int} that provides an
% \ttt{int64} hash result and the geohash in Golang
% Assembly~\cite{Geohash-Golang} that quantizes latitude and longitude to 32-bit
% integers, then bit-interleaves integers to produce a 64-bit value (integer
% geohash), and \ttt{base32} encodes it (with a custom alphabet)
% to produce a geohash string.

Plus Codes can be shortened relative to a reference location, which reduces the number of
digits to use and  we can define the reference
location in a given Context. 
They have similar properties to geohashes and geoprefixes---they
represent areas and the size of the area depends on the code length.

We can store a string version of the geohash and Plus Code in DNS as the names
of an IoT device and enable some geographical/proximity searches using the
\ttt{geohash.org} site or Google Maps (Plus Code required). 

The only constraint of using geo-identifiers for \ttt{DevEUI} is the fact that
\ttt{DevEUI} does not have the EUI-64 format anymore, which may be an obstacle
for some applications. On the other hand, we gain the possibility of linking the
device location with its identifier.

\section{Device Discovery with DNS Queries}\label{sec:impl}

In this section, we discuss the issue of how we can query the DNS system to
discover some properties of IoT devices and find relevant devices corresponding
to those properties. 
We propose to use the defined semantic identifiers and DNS names to offer support for
search and discovery. In constrained environments, providing full-fledged
database search functionality may be difficult. Instead, we propose to take
advantage of the DNS system as the basic functionality for querying and
discovering the semantic properties related to IoT devices. 

\subsection{DNS Service Discovery}\label{sec:dns-sd}

DNS-Based Service Discovery (DNS-SD)~\cite{dnssd} is a functionality of DNS to
discover services in a network.
Information about a given service is stored in the DNS database as an \ttt{SRV} record of
the form:

\begin{footnotesize}
\begin{alltt}
\textbf{<Instance>.<Service>.<Domain> IN SRV <data>}
\end{alltt}
\end{footnotesize}

\noindent
and gives the target host and the preassigned port at which the service instance
can be reached.
The \ttt{TXT} record for the same name gives additional information about this instance
in a structured form using key/value pairs.

A DNS client can discover the list of available instances of a given service
type using a query for a DNS \ttt{PTR} record with a name of the form
\ttt{<Service>.<Domain>} which returns a set of zero or more \ttt{PTR} records giving
the \ttt{<Instance>} names of the services that match the queried \ttt{<Service>}.
Each \ttt{PTR} record is structured as such:

\begin{footnotesize}
\begin{alltt}
\textbf{<Service>.<Dom> IN PTR <Instance>.<Service>.<Dom>}
\end{alltt}
\end{footnotesize}

The \ttt{<Instance>} portion of the \ttt{PTR} data is a user-friendly name consisting of
UTF-8 characters, so rich-text service names and subdomains are allowed and
encouraged, for instance:

\begin{footnotesize}
\begin{alltt}
\textbf{LoRa temp sensor.Room 7.\_iot.\_udp.example.com.}
\end{alltt}
\end{footnotesize}

The \ttt{<Service>} portion of the query consists of a pair of DNS labels,
following the convention already established for \ttt{SRV} records, for instance, the
\ttt{PTR} entry for name \ttt{\_http.\_tcp.local.}:

\begin{footnotesize}
\begin{alltt}
\textbf{\_http.\_tcp.local. PTR web-page.\_http.\_tcp.local.}
\end{alltt}
\end{footnotesize}

\noindent
advertises a ``web-page'' accessible over HTTP/TCP.

We propose to use this mechanism for querying DNS to find devices relevant to
properties or locations expressed in as our semantic names.  For example, the
following query: 

\begin{footnotesize}
\begin{alltt}
\textbf{\_dr5r7p4r.\_iot.\_udp.iot.org IN PTR}
\end{alltt}
\end{footnotesize}

\noindent
would look for IoT devices near the Statue of Liberty.

%\subsection{Use in LoRaWAN Networks}

% LoRa devices are not directly connected to the internet via an IP stack, they
% are in a separate LoRaWAN network, connected to a LoRa Gateway which forwards
% the messages to the Network Server.

\subsection{Structuring Queries as Subdomains}\label{sec:subdomains}

The DNS system stores resource records in a hierarchical tree in which servers can
delegate the management of subdomains. For example, DNS Server A, authoritative
for \ttt{example.fr} can delegate the management of the records for
\ttt{data.example.fr} to DNS Server B. In a similar way, we can delegate the
management of a given geographical region to a specific server whose region is
included in the encompassing region of the delegating domain. For instance, if we want to
delegate the management of the New York area to a city-managed
DNS server, we need to configure a ``New York area'' subdomain and delegate it.

The \ttt{in-addr.arpa} domain, in charge of reverse DNS lookups already uses
this kind of method to delegate the management of an IPv4 address to the owner:
when making a reverse DNS query on \ttt{1.2.3.4}, we query
\ttt{4.3.2.1.in-addr.arpa} that corresponds to the chain of delegations of the
\ttt{1.in-addr.arpa} subdomain for the \ttt{1.0.0.0/8} and so on.

We can use a similar method to split semantic names into multiple subdomains
to easily delegate some properties or locations to other servers. Here is an example for
geo-identifiers: instead of having
to encode all possible geo-identifiers under the \ttt{\_iot.\_udp.iot.org}
domain, we create the \ttt{dr.\_iot.\_udp.iot.org} subdomain and let it handle
all areas with the \ttt{dr} geoprefix (encompassing the east coast of the USA).
This subdomain can delegate subdomains to other servers if needed. 
Similarly, the server in charge of the \ttt{dr} prefix (east coast) can
delegate the \ttt{dr5r} area (encompassing New-York) to another server (a city
managed server for example). Then, the administrator of this server can choose to handle the
\ttt{7p.5r.dr} subdomain, as it represents an area of 600 meters around the
Statue of Liberty.
As a result, instead of querying \ttt{dr5r7p.\_iot.\_udp.iot.org}, we can query for
instance \ttt{7p.5r.dr.\_iot.\_udp.iot.org} and let each subdomain administrator choose if she
wants to delegate some areas to other servers.

There are several ways to split a given semantic name into multiple subdomains
so the user has to know the
number of characters in a given subdomain to use it in a query.
The number of bytes in each subdomain also influences the kind of
queries a user can do. For example, setting $2$ characters per subdomain, like
in 
\ttt{34.12.\_iot.\_udp.iot.org}, makes it impossible to query directly for
devices with the \ttt{123} prefix, so the user has to either query the whole prefix
\ttt{12.\_iot.\_udp.iot.org} and then filter the relevant results, or query
all \ttt{3[0-f].12.\_iot.\_udp.iot.org} domains ($16$ queries). Thus, 
we need to choose the subdomain size carefully. We propose three schemes for
splitting geo-identifiers: a static subdomain length, a dynamic subdomain
length, and multiple subdomain lengths.

\textbf{Static subdomain length.} We set size $S$ for all subdomains.
In this way, the user can split the geo-identifier in several groups of size $S$
(rounded up or down, depending on the preference of a query on the encompassing zone and
then filtering, or making multiple sub-queries) without any additional knowledge. 
The drawback is the lack of flexibility and the arbitrary choice of $S$ that may
be suitable for a given area but not for another one.

\textbf{Dynamic subdomain length.} Each domain has a \ttt{TXT} record that gives the
size of the subdomains related to a given area. For example, \ttt{12.\_iot.\_udp.iot.org
IN TXT "len=3"} informs the user that under the \ttt{12} subdomain, each
subdomain has length $3$, so one can query \ttt{345.12.\_iot.\_udp.iot.org}.
This scheme supports the right subdomain length for each region: in a dense
area where we need multiple precise subdomain delegations, we can set a small
length to obtain precise subdivision and in sparse areas where we do not need
small subdivisions (seas, fields), we can use a larger length (3 or 4). 
The scheme supports multiple subdomain lengths in
the same query like in \ttt{6.345.12.\_iot.\_udp.iot.org} as each subdomain
can set its size. 
The drawback is the need for recursively 
querying different subdomains for their \ttt{TXT} records to know the length of
each field before splitting the query the right way. 

\textbf{Multiple subdomain lengths.} There are multiple ways to get to a given
subdomain, so multiple ways of splitting the geo-identifier are possible and valid. For
example, both \ttt{345.12.\_iot.\_udp.iot.org} and
\ttt{5.34.12.\_iot.\_udp.iot.org} are valid and point to the same area. 
In this way, the users do not have to query for \ttt{TXT} records and can split their
queries as they want. 
However, it may be hard to encode all ways of splitting the geo-identifier into
subdomains in a resource record. 

We can simplify this method with \ttt{CNAME} records,
the same way the \ttt{in-addr.arpa} domain handles the delegation of
subnetworks with arbitrary
size\footnote{\url{https://tools.ietf.org/html/rfc2317}} by defining multiple
\ttt{CNAME} records.
For example, if two different servers need to handle the
prefixes \ttt{12a} and \ttt{12b} but
the \ttt{12.\_iot.\_udp.iot.org} domain only defines subdomains of length $2$, we
can insert the following records:

\begin{footnotesize}
\begin{alltt}
  \textbf{a NS server.handling.a.12.area}
  \textbf{a0 CNAME 0.a.12.\_iot.\_udp.iot.org}
  \textbf{a1 CNAME 1.a.12.\_iot.\_udp.iot.org}
  \textbf{a2 CNAME 2.a.12.\_iot.\_udp.iot.org}
  \textbf{\ldots}
  \textbf{af CNAME f.a.12.\_iot.\_udp.iot.org}
\end{alltt}
\end{footnotesize}

%We can do the same for all $16$ \ttt{bX.12} records. 
We can apply the same approach to all $16$ \ttt{bX.12} records. 
Once the \ttt{CNAME} records
are created, a user querying \ttt{a2.12.\_iot.\_udp.iot.org} will be redirected to
\ttt{2.a.12.\_iot.\_udp.iot.org}, so she will try to resolve the \ttt{a.12}
part and will receive an \ttt{NS} entry pointing to the server in charge of the \ttt{12a}
area. Therefore, with these records, the user does not have to know how the
delegation in the \ttt{12} area works, the query does not change from her point
of view, but with \ttt{CNAME} and \ttt{NS} records, we can transparently delegate
parts of the subdomain. Moreover, this method allows for easy modification of the
server in charge of (authoritative for) \ttt{a.12} because changing the \ttt{NS} entry is easy and the \ttt{CNAME}
records remain the same. 
The method may generate many \ttt{CNAME} entries but they are simple to generate
automatically and do not need to change often.

We can use the subdomain splitting for different contexts like for logical
localizations. In this particular case, we can easily encode the properties in
different subdomains because they are naturally ordered (a room on a given floor
in a given building). For example, if the Context for Logical Localization is
\ttt{2}, the position of a
device in Building 1 on Floor 5 in Room 56 is as follows (with \ttt{base32} geohash
in the parenthesis):

%\vspace{1em}
\begin{itemize}
  \item Context-2: $2$ - (\ttt{2})
  \item Building: $1$ - (\ttt{1})
  \item Floor: $5$ - (\ttt{5})
  \item Room: $56$ - (\ttt{1s})
\end{itemize}

So, to get the sensors in this room, we send the following query:
\begin{footnotesize}
\begin{alltt}
\textbf{\_1s.\_5.\_1.\_2.\_iot.\_udp.iot.org IN PTR}
\end{alltt}
\end{footnotesize}

\subsection{Use of \ttt{AXFR} for the Result Set}

%Once we know which domain to query, we need a way to obtain all devices and zone
%information related to a given domain (e.g., temperature measured by all sensor
%in a given logical or geographic location). We can use the DNS-SD protocol
%described previously to express precise queries, however, DNS defines a means
%for getting all records in a zone with the DNS Zone Transfer Protocol
%(AXFR)\footnote{\url{https://tools.ietf.org/html/rfc5936}}. 
Another way of obtaining the result set from a DNS server is to use the DNS Zone
Transfer Protocol (\ttt{AXFR})~\cite{axfr} %\footnote{\url{https://tools.ietf.org/html/rfc5936}}
that returns all records in a zone. When a client sends an \ttt{AXFR} query message to
an authoritative server, it answers with all resource records stored in the
zone. 
%Not all servers answer an AXFR query, as it requires good bandwidth
%but we can take advantage of this feature to return the results of a query on subdomains describing a geographical
%area small enough so that the number of devices is reasonable. 
This feature can be used to return the results of a query on subdomains
describing a property or a geographical area of the interest. For instance, to get all devices
and the corresponding data stored in the zone in \ttt{123456}, the user can use
the following command:

\begin{footnotesize}
\begin{alltt}
\textbf{dig @server AXFR 56.34.12.\_iot.\_udp.iot.fr}
\end{alltt}
\end{footnotesize}

\section{Prototype Implementation of Semantic Discovery}
\label{sec:proto}

We have implemented two prototypes for geo-identifiers of LoRa devices. Their
extension to consider other types of semantic names is undergoing. 
The prototypes for geo-identifiers are available to the public 
to encourage reproducibility.
We show below some examples of their utilization with commands using the \ttt{dig} tool. 
These prototypes only consider geohashes encoded in the domain
name without the use of the Context described in
Section~\ref{sec:encoding}.

The first prototype,\footnote{\url{https://github.com/dsg-unipr/geo-dns}} which takes
advantage of the Node-based \emph{dns2} module~\cite{dns2} and the Redis
in-memory database,\footnote{\url{https://redis.io}} allowed us to quickly deploy
and test the concepts based on hardcoded data.

\begin{figure}[t]%placement
   \centering
\includegraphics[width=0.8\linewidth]{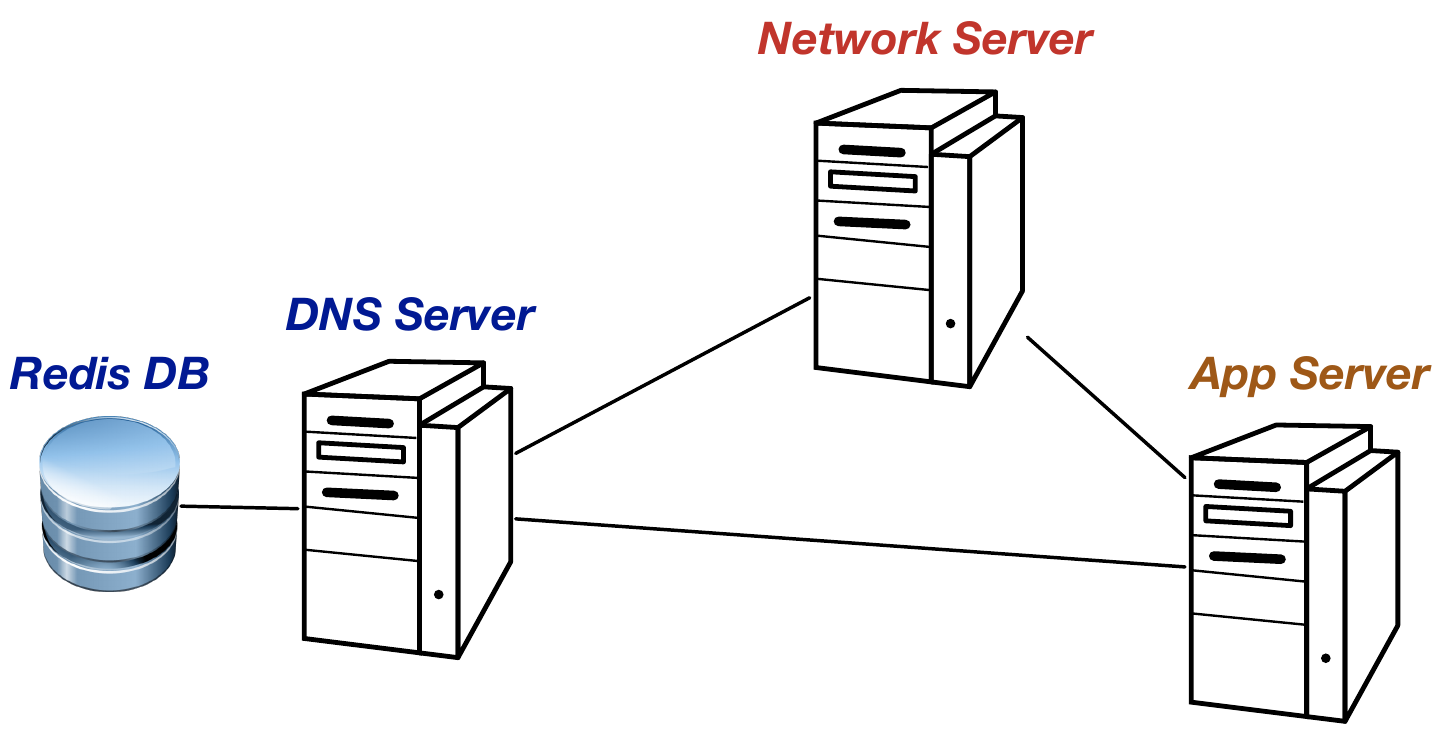}\\[-1ex]
   \caption{Prototype for LoRa geo-identifiers based on DNS-SD.}
   \label{fig:proto}
\vspace{-2ex}
\end{figure}

The second prototype\footnote{\url{https://github.com/fabrizior/coredns}} uses
the CoreDNS DNS server written in Go.\footnote{\url{https://coredns.io}} CoreDNS
is highly flexible thanks to plugins that perform different functions: DNS,
Kubernetes service discovery, Prometheus metrics, rewriting queries, and many
more. We have modified the \textit{file} plugin that enables serving zone data
from an RFC 1035-style master file. 

In our prototypes, Applications or Network Servers that want to discover the location
of LoRa devices can query a DNS server to find the devices matching some
criteria based on their location. 
In other types of networks, devices themselves can directly query a DNS server. 

In a LoRa network with geo-identifiers, when registering a device, the Network or Join
Server register several records in the DNS database. First, an \ttt{SRV} record
giving the domain and ports where the Network Server managing a given device can be
queried. Then, \ttt{PTR} records that allows finding the device based on its
geo-identifier or name:

\begin{footnotesize}
\begin{alltt}
\textbf{<name>.\_iot.\_udp.<Domain> IN SRV <port> <domain>}
\textbf{\_<geo-i>.\_iot.\_udp.<Domain> IN PTR <name>}
\end{alltt}
\end{footnotesize}

\ttt{name} being the semantic name like described in Section~\ref{sec:encoding}.
This name of the given domain is unique and describes the properties of the
device. \ttt{geo-i} is the geo-identifier of the device encoded in multiple
subdomains as described in Section~\ref{sec:subdomains}.
When an application needs to find all devices in a given area, it can
query DNS for all devices in the matching subdomain by sending a query like:

\begin{footnotesize}
\begin{alltt}
\textbf{\_<geo-i>.\_iot.\_udp.<Domain> IN PTR},
\end{alltt}
\end{footnotesize}

where \ttt{geo-i} can be split into multiple subdomains if needed.

The DNS server answers with the list of all \ttt{PTR} records in the queried
subdomain, and therefore, in the represented area. Each \ttt{PTR} record gives the
semantic name of a device in the area. Once the application knows the name of the devices in
the area, it can query the DNS server for an \ttt{SRV} record with the different
semantic name and get the Network Server managing the devices.

We have implemented this method in our prototypes and both of them can be queried
with the \ttt{dig} tool.\footnote{The address 127.0.0.1 used in the following
examples should be replaced with the actual IP address of the DNS server.} Upon
receiving a \ttt{PTR} query for a specific \ttt{<Service>}, the server returns all
instances of that service type in the subdomain:

\begin{scriptsize}
\begin{alltt}
\textbf{\# dig @127.0.0.1 -p 53 \_dr.\_iot.\_udp -t PTR}
\textbf{;; QUESTION SECTION:}
\textbf{;\_dr.\_iot.\_udp.			IN	PTR}

\textbf{;; ANSWER SECTION:}
\textbf{\_dr.\_iot.\_udp.		100	IN	PTR	humidity.dr12.\_iot.\_udp.}
\textbf{\_dr.\_iot.\_udp.		100	IN	PTR	temperature.dr34.\_iot.\_udp.}
\textbf{\_dr.\_iot.\_udp.		100	IN	PTR	temperature.dr56.\_iot.\_udp.}
\end{alltt}
\end{scriptsize}

Then, once the application has obtained the semantic name of the device (for example,
\ttt{temperature.dr56}), it can query the server for an \ttt{SRV} record with this
name, which will contain the domain and ports at which access the device. 
It can also ask for \ttt{TXT} records to get additional data about
the device. For example, still using \ttt{dig}:

\begin{scriptsize}
\begin{alltt}
\textbf{\# dig @127.0.0.1 -p 53 temperature.dr56.\_iot.\_udp -t ALL}
\textbf{;; QUESTION SECTION:}
\textbf{;temperature.dr56.\_iot.\_udp.		IN	ALL}

\textbf{;; ANSWER SECTION:}
\textbf{temperature.dr56.\_iot.\_udp.	100	IN	SRV	10 20 8080 dr56.unipr.it.}
\textbf{temperature.dr56.\_iot.\_udp.	100	IN	TXT	"temperature=14"}
\end{alltt}
\end{scriptsize}

Finally, when an A query for the \ttt{<Domain>} managing a device is received,
the server returns the IP address of the Network Server the LoRa device is
associated with. For example, if the following \ttt{dig} command is executed: 

\begin{scriptsize}
\begin{alltt}
\textbf{\# dig @127.0.0.1 -p 53 dr56.unipr.it -t A}
\textbf{;; QUESTION SECTION:}
\textbf{;dr56.unipr.it.		IN	A}
\end{alltt}
\end{scriptsize}

we obtain the following answer: 
\begin{scriptsize}
\begin{alltt}
\textbf{;; ANSWER SECTION:}
\textbf{dr56.unipr.it.	100	IN	A	160.78.28.203}
\end{alltt}
\end{scriptsize}

\section{DNS as a Source of IoT Data}\label{sec:dns-data}

In the previous sections, we have presented the schemes for encoding device
properties in domain names to discover devices by querying the DNS
infrastructure. Once the user discovers some relevant devices, she still needs to
contact them with different protocols to obtain data or set up data delivery
process with the COAP Observe option for instance. 
We can also take advantage of the DNS infrastructure as a public store for IoT
data in a similar way to the Cloud. 
Many IoT applications store data in the Cloud for further processing and access
by clients. 

The idea of DNS as a source of IoT data is to use a \ttt{TXT} record associated with a
name of an IoT device to store its data so that a large number of users can
access the data in DNS instead of getting them directly from the device. 
As a \ttt{TXT} record linked to a domain is usually already filled with human readable
data related to the domain, we can add dynamically created records. 
Once the IoT data is stored in the \ttt{TXT} record, users will benefit from the DNS caching
infrastructure efficient dissemination: recursive resolvers will cache its
content and keep the data until the time-to-live (\ttt{ttl}) of the record
expires. Then, the recursive resolvers will query the authoritative server to
get the new record and the updated data from the device. With this method, the
end users do not need to know what kind of a protocol should be used to contact
the device, as data are stored in a standard \ttt{TXT} record and no direct contact 
between the user and the device is required. 

\subsection{Encoding Data in \ttt{TXT} Records} 
RFC 6950\footnote{\url{https://tools.ietf.org/html/rfc6950}} describes under what 
conditions an application can use DNS to store data and provides several
recommendations and warnings indicated by other RFCs. 
RFC 1464\footnote{\url{https://tools.ietf.org/html/rfc1464}} formalized the
\ttt{<key>=<value>} format for storing data in \ttt{TXT} records, so in the case of
the example of a temperature sensor, the DNS entry should be \ttt{<domain> IN
  <ttl> TXT "temperature=14"}.

Not all types of data should be placed in DNS: records with a large size can be
used by attackers as an amplifier to generate a lot of
traffic~\cite{Rossow14amplificationhell} (this is why 
\ttt{.com} records are limited to 1460 bytes). Therefore, this solution may not be
suitable for all kinds of sensors. For example, a device taking periodic $512
\times 512$ pictures would generate data that should not be put on DNS, instead,
the user will have to find a way to contact the device directly, or its Network
Server to get the data from a suitable source.

\subsection{Updating Data in \ttt{TXT} Records} 

To keep data in the DNS record updated, there should be a process or an entity
that gets the data from the device and updates the corresponding \ttt{TXT} record. For
non-constrained devices, an IoT device could update its own record, but for most
constrained devices, this kind of operation may be too costly, so another entity
can update data. 
For LoRa networks, all data from the devices go through the Network Server.
As this server is not constrained, it can update the \ttt{TXT} record on behalf of the
device using, for example, secure Dynamic DNS Update protocol
extension~\cite{zone} 
%\footnote{\url{https://tools.ietf.org/html/rfc2136}}\textsuperscript{,}\footnote{\url{https://tools.ietf.org/html/rfc3007}}
and a standard Unix \ttt{nsupdate} command to insert the new values in the zone file of the 
authoritative DNS server.

Because the data are not updated in real time, it is important to choose a suitable
\ttt{ttl} value of the \ttt{TXT} record, so that the data are ``marked'' as out of date when new
values are available. The \ttt{ttl} value must take into account the frequency at which the Network
Server retrieves the new data from the device and updates the corresponding DNS record.
For example, if the Network Server retrieves the temperature data and dynamically updates 
\ttt{TXT} records every hour, then the \ttt{ttl} value should be ``synchronized'' and also 
set to one hour so that the information stored in caches of local DNS resolvers, which 
request the data on behalf of local clients, is also up to date. 

We can also use the Incremental Transfer mechanism 
(\ttt{IXFR})\footnote{\url{https://tools.ietf.org/html/rfc1995}} designed to
transfer only a modified part of a zone, for example,  the updated \ttt{TXT}
records with the changed temperature. Each time the zone is dynamically updated
by, for example, the Network Server, the serial number of its zone is increased.
Therefore, after the initial \ttt{AXFR} transfer, the client should keep record of the
Start of Authority (\ttt{SOA}) serial number of the transferred zone. Next, the client
can send an \ttt{IXFR} request with the registered version number so that the
authoritative name server responds only with the deleted and added resource
records since the version known by the \ttt{IXFR} client up to the current version of
the zone stored by the authoritative server. For example, to get new data
related to the \ttt{123456} location, the client can use the following command:

\begin{footnotesize}
\begin{alltt}
\textbf{dig @server IXFR=[old-ser] 56.34.12.\_iot.\_udp.iot.fr}
\end{alltt}
\end{footnotesize}

%\balance

\section{Conclusion}\label{sec:conclusion}
\label{sec:conclusion}

In this paper, we have proposed a scheme for representing semantic metadata of IoT
devices in compact identifiers  and DNS names to enable simple discovery and search
with standard DNS servers.
Our scheme defines a binary identifier as a sequence of bits composed of a
Context and several bits of fields encoding semantic properties specific to the Context. The bit
string is then encoded as a character string, stored in DNS.
In this way, we may take advantage of the DNS system 
as the basic functionality for querying and discovery of semantic
properties related to IoT devices. 

We have defined specific Contexts for hierarchical properties as well as logical 
and geographic locations. 
For this last part, we have developed two prototypes that manage geo-identifiers
in LoRa networks to show that the proposed scheme can take advantage of the
standard DNS infrastructure.  
%We may think about Geo-LoRa as a complementary method to other existing geolocation schemes such as TDoA.
%The encoding of LoRa identifiers on 64 bits has the property of progressive
%resolution---skipping the least significant bits results in defining larger areas,
%thus enabling queries over regions that return all nodes located in a given
%place. Finally, we presented two Geo-LoRa prototypes, based on high performance DNS servers.

Regarding future work, we will thoroughly assess the proposed approach using
deployed LoRaWAN devices. Furthermore, we plan to further investigate the idea
of DNS as a source of IoT data, with particular attention to the problem of
getting data from the devices and updating the \ttt{TXT} records. 

%\section*{Acknowledgments}
%\todo{Write the acknowledgments}
%\vspace{-0.5em}
\section*{Acknowledgments}

%\small
This work has been partially supported by the French Ministry of Research
projects PERSYVAL-Lab under contract ANR-11-LABX-0025-01 and DiNS under contract
ANR-19-CE25-0009-01.

\bibliographystyle{IEEEtran}
\bibliography{all,lora,lora2,dt,lora2019_v2,geo,dinas,sensor,biblio,dns-dane}

\end{document}